\documentclass[aps,showpacs,twocolumn,superscriptaddress]{revtex4}
\usepackage[dvips]{graphicx}
\usepackage{amsmath}
\usepackage{color}

\begin{document}

\title{Temperature and Frequency Dependent Mean Free Paths of Renormalized Phonons in Nonlinear Lattices}

\author{Nianbei Li}
\affiliation{Center for Phononics and
Thermal Energy Science and School of Physics Science and
Engineering, Tongji University, 200092 Shanghai, China}

\author{Junjie Liu}
\affiliation{State Key Laboratory of Surface Physics and Department of Physics, Fudan University, 200433 Shanghai, China}

\author{Changqin Wu}
\affiliation{State Key Laboratory of Surface Physics and Department of Physics, Fudan University, 200433 Shanghai, China}

\author{Baowen Li}
\email{phononics@tongji.edu.cn} \affiliation{Department of Mechanical Engineering, University of Colorado Boulder, CO 80309, USA}

\begin{abstract}
In the regime of strong nonlinearity, the validity of conventional perturbation based phonon transport theories is questionable. In particular, the renormalized phonons instead of phonons are responsible for heat transport in nonlinear lattices. In this work, we directly study the temperature and frequency dependent Mean Free Path (MFP) of renormalized phonons with the newly developed numerical tuning fork method. The typical 1D nonlinear lattices such as Fermi-Pasta-Ulam $\beta$ (FPU-$\beta$) lattice and $\phi^4$ lattice are investigated in details. It is found that the MFPs are inversely proportional to the frequencies of renormalized phonons rather than the square of phonon frequencies predicted by existing phonon scattering theory.
\end{abstract}
\pacs{05.60.-k,44.10.+i,05.45.-a}

\maketitle
Phonon is a basic concept in solid state physics describing the collective motions of lattice vibrations. The phonon description is rigorously precise only for Harmonic lattices. For nonlinear lattices especially when nonlinearity cannot be treated as a small perturbation, the concept of renormalized phonons emerges and theoretical efforts have been devoted to describe these novel collective motions. The renormalized phonons are discovered by different groups independently in various research areas ranging from lattice vibrations\cite{Alabiso1995,Alabiso2001}, heat conduction\cite{Lepri1998}, field thermalization\cite{Boyanovsky2004} and nonlinear waves\cite{Gershgorin2005,Gershgorin2007}. The theoretical predictions of the dispersion relations of renormalized phonons from these different approaches are found to be approximate but not exact to each other\cite{Nianbei2006,Dahai2008}. This puzzle is solved by a recent variational approach which unifies the renormalized phonon theory by applying suitable approximations in a systematical way\cite{Junjie2015}. The existence of renormalized phonons in general 1D nonlinear lattices has been verified numerically in computer simulations\cite{Alabiso1995,Alabiso2001,Gershgorin2005,Gershgorin2007,Nianbei2006,Nianbei2013,Linlin2014}. The role of renormalized phonons as the heat energy carriers has also been proposed and testified by numerical simulations\cite{Nianbei2007,Nianbei2009,Nianbei2010,Nianbei2012,Nianbei2013,Linlin2014}.

In the regime of strong nonlinearity, the validity of conventional perturbation based phonon transport theories is questionable. The phenomenological effective phonon theory\cite{Nianbei2006,Nianbei2007,Nianbei2012,Nianbei2013,Linlin2014} is developed within the framework of renormalized phonons dedicated to the explanations of temperature dependence of thermal conductivities for nonlinear lattices. This theory can predict the actual exponents of the power-law dependence of thermal conductivities as the function of temperature for typical 1D nonlinear lattices. For example, the effective phonon theory predicts the temperature dependent thermal conductivities $\kappa(T)$ of 1D FPU-$\beta$ lattice are inversely proportional to temperature as $\kappa(T)\propto T^{-1}$ at low temperature region and proportional to the quartic root of temperature as $\kappa(T)\propto T^{1/4}$ at high temperature region\cite{Nianbei2007,Nianbei2012}. It also predicts the temperature dependence is $\kappa(T)\propto T^{1/2-1/n}$ for $H_n$ lattices\cite{Nianbei2012}. For lattices with on-site potentials, the effective phonon theory predicts for 1D $\phi^4$ lattice the temperature dependence is $\kappa(T)\propto T^{-4/3}$\cite{Nianbei2013}. For general nonlinear Klein-Gordon lattices where $\phi^4$ lattice is a special example of $n=4$, this theory predicts the general temperature dependence as $\kappa(T)\propto T^{4(2-n)/(n+2)}$\cite{Linlin2014}. Extensive numerical simulations have verified these predictions quantitatively and consistently for FPU-$\beta$ lattice\cite{Nianbei2007,Nianbei2012,Aoki2001prl}, $H_n$ lattices with $n=3,4,5$\cite{Nianbei2012}, nonlinear Klein-Gordon lattices with $n=1.25,1.5,1.75,2.5,3,3.5,4$\cite{Nianbei2013,Linlin2014,Aoki2000pla}. It should be emphasized that these theoretical predictions are derived from effective phonon theory without any fitting parameter.

In determining the thermal conductivities, the most important information is the temperature and frequency dependence of MFPs of energy carriers. The newly developed tuning fork method proposed a numerical method to calculate the MFPs for every phonon mode in a direct way \cite{Sha2014prb}. This tuning fork method enables us to testify the conjecture of temperature and frequency dependence of MFPs made in effective phonon theory \cite{Nianbei2012,Nianbei2013}: the renormalized phonon's MFP is inversely proportional to the renormalized phonon frequency $\hat{\omega}_k$ and the nonlinearity strength $\varepsilon$ as
\begin{equation}\label{conjecture}
\hat{l}_k\propto\frac{\hat{v}_k}{\varepsilon\hat{\omega}_k},
\end{equation}
where $\hat{v}_k=\partial{\hat{\omega}_k}/\partial{k}$ is the group velocity of renormalized phonons. The dimensionless nonlinearity strength $\varepsilon$ is defined as the ratio of ensemble averaged nonlinear potential energy $E_n$ and total potential energy $E_t=E_l+E_n$ to be $\varepsilon=E_n/(E_l+E_n)$ where $E_l$ is the ensemble averaged linear potential energy.

In this work, we will directly test the validity of this conjecture by using the newly developed tuning fork method\cite{Sha2014prb}. The temperature and frequency dependent MFPs of renormalized phonons will be calculated and compared with the conjecture for typical 1D FPU-$\beta$, $H_4$ and $\phi^4$ lattices. The good agreement between numerical results and the conjecture indicates that the MFP of renormalized phonons is indeed inversely proportional to the renormalized phonon's frequency and the nonlinearity strength, which is beyond the scope of any conventional perturbative phonon transport theory.

\begin{figure}
\includegraphics[width=0.8\columnwidth]{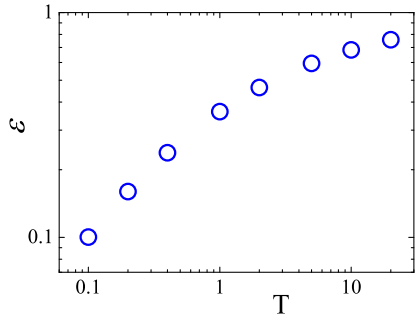}
\vspace{-0.5cm} \caption{\label{fig:epsilon-fpub} (color online). The dimensionless nonlinearity strength $\varepsilon$ as the function of temperature for the FPU-$\beta$ lattice. The $\varepsilon$ increases linearly with temperature as $\varepsilon\propto T$ in the low temperature regime and saturates to maximum value of $1$ as $\varepsilon\rightarrow 1$ in the high temperature limit. These values are evaluated from Eq. (\ref{epsilon-fpub}) and these temperature points will be used to calculate the MFPs of renormalized phonons as in Fig. \ref{fig:fpub}.}
\end{figure}

We consider the 1D nonlinear lattices of $N$ atoms with the general Hamiltonian
\begin{equation}
H=\sum_{i}\left[\frac{p^2_i}{2}+V(q_{i+1}-q_{i})+U(q_i)\right]
\end{equation}
where $p_i$ and $q_i$ denote the momentum and displacement of the $i$-th atom, respectively. For simplicity, periodic boundary conditions of $q_i=q_{N+i}$ and dimensionless units have been used. The $V(x)$ represents the inter-atom potential energy where only nearest neighbor interaction has been considered and the $U(x)$ is the on-site potential energy. In this work, three typical 1D nonlinear lattices either with or without on-site potentials will be investigated, namely, the FPU-$\beta$, $H_4$ and $\phi^4$ lattices. The combination of $V(x)=x^2/2+x^4/4$ and $U(x)=0$ describes the FPU-$\beta$ lattice while the combination of $V(x)=x^4/4$ and $U(x)=0$ denotes the $H_4$ lattice. For the $\phi^4$ lattice, the potential energy takes the form of $V(x)=x^2/2$ and $U(x)=x^4/4$.

The renormalized phonon frequency $\hat{\omega}_k$ can be expressed in a general form as $\hat{\omega}_k=\sqrt{\alpha\omega^2_k+\gamma}$ where $\omega_k=2\sin{\frac{k}{2}}, \,\, -\pi<k\leq \pi$ is the familiar phonon frequency of the Harmonic lattice\cite{Nianbei2012}. The renormalization coefficient $\alpha$ is determined only by the inter-atom potential energy $V(x)$ while the coefficient $\gamma$ depends only on the on-site potential energy $U(x)$. For Harmonic lattice, $\gamma=0$ and $\alpha=1$ ensure that $\hat{\omega}_k=\omega_k$ is recovered. The nonlinearity strength $\varepsilon$ defined as $\varepsilon=E_n/(E_l+E_n)$ is a dimensionless unit. For simplicity, we only consider lattices with hard potential with which the $\varepsilon$ will not be ill-defined to be negative value. For the Harmonic lattice, the nonlinearity strength $\varepsilon$ equals to zero which should be expected since Harmonic lattice is a linear system. The existence of renormalized phonons in these lattices have been verified by numerical simulations\cite{Alabiso1995,Alabiso2001,Gershgorin2005,Gershgorin2007,Nianbei2006,Nianbei2013,Linlin2014,Junjie2015}.

\textbf{FPU-$\beta$ lattice.} For lattices without on-site potential, the renormalization coefficient $\gamma$ is zero. The expression of renormalized phonon frequency and its group velocity can be simplified as
\begin{equation}\label{fpub-omega}
\hat{\omega}_k=\sqrt{\alpha}\omega_k,\,\,\, \hat{v}_k=\sqrt{\alpha}v_k
\end{equation}
where $v_k=\cos{\frac{k}{2}}$ is the phonon group velocity for the Harmonic lattice. Here the renormalization coefficient $\alpha$ can be analytically obtained as $\alpha=1+\int^{\infty}_{0}x^4 e^{-(x^2/2+x^4/4)/T}dx/\int^{\infty}_{0}x^2 e^{-(x^2/2+x^4/4)/T}dx$ which is only temperature or equivalently nonlinearity dependent \cite{Nianbei2010}.

\begin{figure}
\includegraphics[width=0.8\columnwidth]{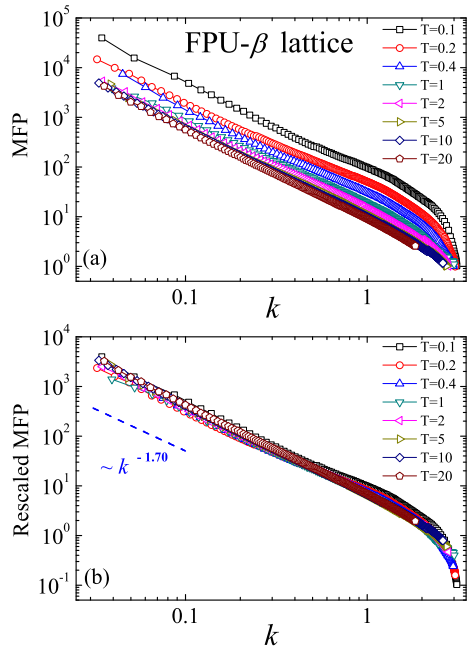}
\vspace{-0.5cm} \caption{\label{fig:fpub} (color online). (a) MFP $\hat{l}_k$ of renormalized phonons as the function of $k$ for FPU-$\beta$ lattice at different temperatures. The larger the temperature, the larger the nonlinearity and the shorter the MFPs as revealed by Eq. (\ref{mfp-fpub}). (b) Rescaled MFP $(\hat{l}_k\cdot\varepsilon)$ as the function of $k$ for FPU-$\beta$ lattice at different temperatures. The dashed line of $k^{-1.70}$ is the guide for the eye. The calculations are performed on lattices with size $N=1024$ for $T<5$ and $N=2048$ for $T\ge 5$.}
\end{figure}

For the FPU-$\beta$ lattice, the nonlinearity strength $\varepsilon$ can be expressed as $\varepsilon=\left<x^4/4\right>/(\left<x^2/2\right>+\left<x^4/4\right>)$ where $\left<\cdot\right>$ means ensemble average. It can further be analytically resolved as
\begin{equation}\label{epsilon-fpub}
\varepsilon=\frac{1}{2\frac{\int^{\infty}_{0}x^2 e^{-(x^2/2+x^4/4)/T}dx}{\int^{\infty}_{0}x^4 e^{-(x^2/2+x^4/4)/T}dx}+1}
\end{equation}
This temperature dependence of $\varepsilon$ for the 1D FPU-$\beta$ lattice is plotted in Fig. \ref{fig:epsilon-fpub}. At low temperature limit, $\varepsilon$ increases with temperature linearly as $\varepsilon\propto T$, while $\varepsilon$ approaches to maximum value $1$ at the high temperature limit. The temperatures of these data points will be the temperatures at which the MFPs of renormalized phonons are calculated. It can be seen that the range of $\varepsilon$ values considered here is about one order of magnitude.

According to the conjecture of Eq. (\ref{conjecture}), the MFPs of renormalized phonons for the FPU-$\beta$ lattice is $\hat{l}_k\propto v_k/(\varepsilon\omega_k)$ since the coefficient $\alpha$ is cancelled for $\hat{v}_k$ and $\hat{\omega}_k$. If we only consider the temperature dependence of MFPs $\hat{l}_k$, we have
\begin{equation}\label{mfp-fpub}
\hat{l}_k\propto \frac{1}{\varepsilon},
\end{equation}
which is a global effect. It says that for the temperature effect, the MFP of every renormalized phonon is inversely proportional to the nonlinearity strength $\varepsilon$.

\begin{figure}
\includegraphics[width=0.8\columnwidth]{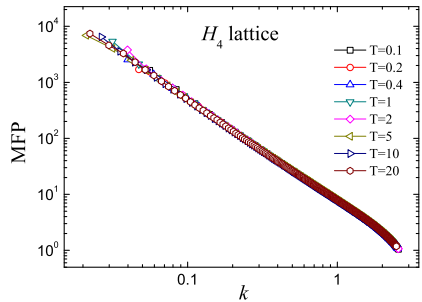}
\vspace{-0.5cm} \caption{\label{fig:h4} (color online). MFPs $\hat{l}_k$ of renormalized phonons  as the function of $k$ for $H_4$ lattice at different temperatures. The MFPs are found to be totally independent of temperatures which is exactly what the conjecture predicts. The calculations are performed on lattices with size $N=1024$ for $T<5$ and $N=2048$ for $T\ge 5$.}
\end{figure}

In Fig. \ref{fig:fpub}(a), the MFPs of renormalized phonons are plotted for different temperatures from $T=0.1$ to $T=20$. The lower the temperature, the longer the MFPs due to its small nonlinearity strength $\varepsilon$ as indicated from Eq. (\ref{mfp-fpub}). In order to quantitatively verify the dependence of Eq. (\ref{mfp-fpub}), we plot the rescaled MFPs of $\hat{l}_k \varepsilon$ as the function of temperature in Fig. \ref{fig:fpub}(b). It can be seen that all the rescaled MFPs $\hat{l}_k\varepsilon$ at different temperatures collapse into a single curve, which is a clear verification of the Eq. (\ref{mfp-fpub}).

It should be mentioned that the numerically calculated MFPs diverges as $\hat{l}_k\propto k^{-1.70}$ at low frequency limit as $k\rightarrow 0$ as observed in Ref. \cite{Sha2014prb}, which finally gives rise to a divergent thermal conductivity as $\kappa\propto N^{0.41}$ known as the anomalous heat conduction\cite{Lepri1997prl,Lepri2003pr,Dhar2008ap,Liu2013epjb,Lepri1998epl,Hatano1999pre,Alonso1999prl,Narayan2002prl,Zhang2002pre,Pereverzev2003pre,Wang2004prl,Zhao2005prl,Zhang2005jcp,Cipriani2005prl,Pereira2006prl,
Delfini2006pre,Basile2006prl,Zhao2006prl,Dhar2007prl,Henry2009prb,Wang2011epl,Wang2012pre,Beijeren2012prl,Dhar2013pre,Mendl2013prl,Pereira2013pre,Spohn2014jsp,Liu2014prl,
Savin2014pre,Wang2015pre,Chang2008prl,Xu2014nc,
Meier2014prl}. The conjecture of $\hat{l}_k$ in the Eq. (\ref{conjecture}) can only provide a $\hat{l}_k\propto k^{-1}$ dependence. However, this extra $k^{-0.70}$ dependence cannot come from the renormalized phonon frequency $\hat{\omega}_k$ which will bring additional temperature dependence to MFPs $\hat{l}_k$. This property is unique for momentum-conserving lattices and the reason for this is still an open issue.

\textbf{$H_4$ lattice.} As a momentum-conserving lattice, the $H_4$ lattice has the same expression for renormalized phonon frequency and its group velocity as in Eq. (\ref{fpub-omega}). But the renormalization coefficient $\alpha$ has a different expression as $\alpha=2\Gamma(5/4)/\Gamma(3/4)T^{1/2}$ \cite{Nianbei2010}.

The $H_4$ lattice is the high temperature limit of the FPU-$\beta$ lattice and is a pure nonlinear lattice untreatable for any perturbative phonon transport theory. There is no quadratic term in the inter-atom potential energy of $V(x)=x^4/4$. By definition, the nonlinearity strength $\varepsilon$ achieves the maximum value as $\varepsilon=1$ for all temperatures. Same as the case for the FPU-$\beta$ lattice, the MFPs $\hat{l}_k$ of $H_4$ lattice shares the same temperature dependence as $\hat{l}_k\propto \varepsilon^{-1}$. However, since $\varepsilon=1$ at any temperature for the special $H_4$ lattice, the MFPs
\begin{equation}\label{mfp-h4}
\hat{l}_k\propto \text{constant},
\end{equation}
which does not depend on temperature at all!

In Fig. \ref{fig:h4}, the MFPs of renormalized phonons in the $H_4$ lattice are plotted for temperatures from $T=0.1$ to $T=20$. All the MFPs collapse into a single curve showing no sign of temperature dependence. The remarkable feature of Eq. (\ref{mfp-h4}) is clearly verified.

\begin{figure}
\includegraphics[width=0.8\columnwidth]{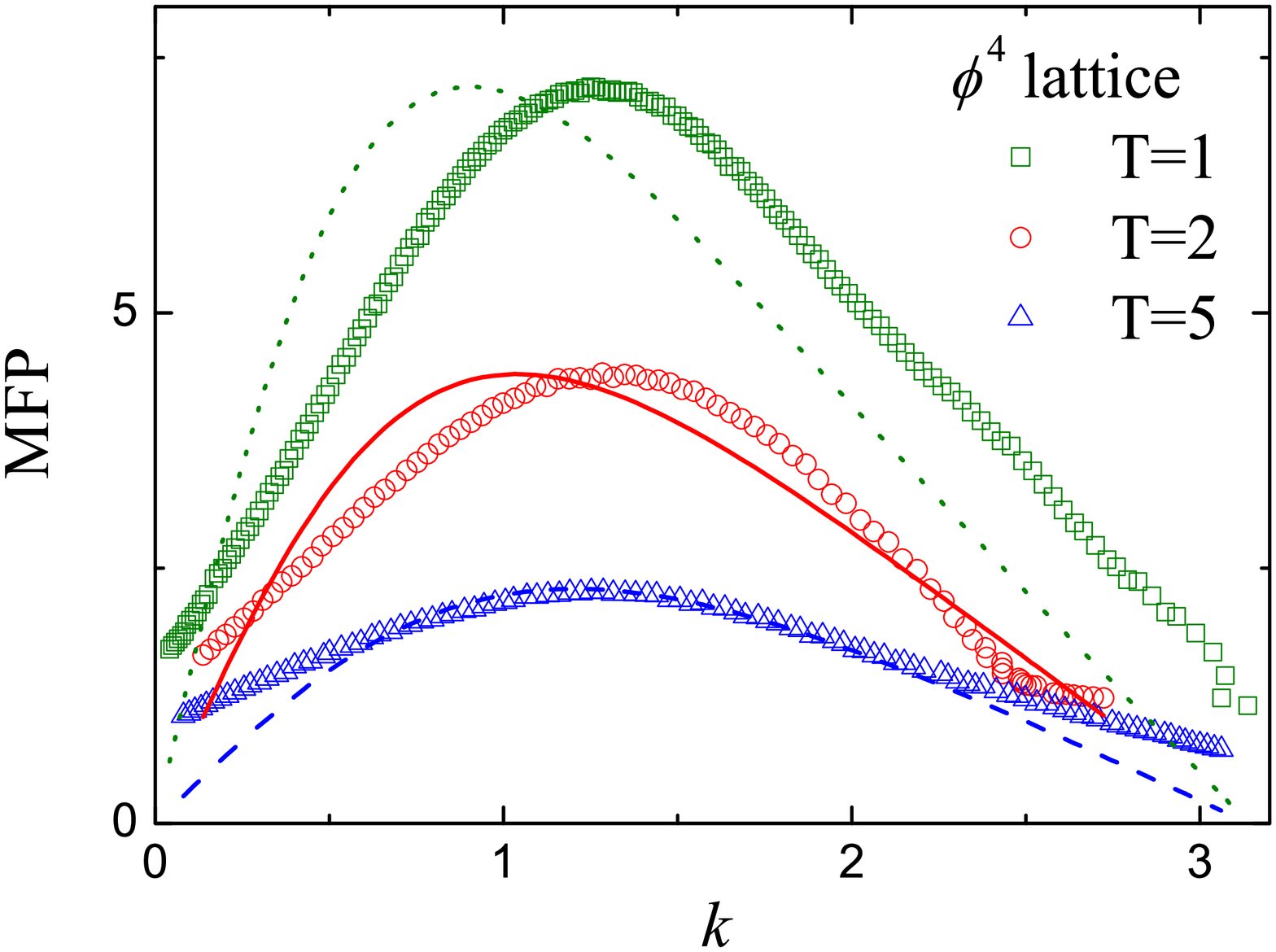}
\vspace{-0.5cm} \caption{\label{fig:phi4-k} (color online). MFPs $\hat{l}_k$ of renormalized phonons  as the function of $k$ for $\phi^4$ lattice at different temperatures at $T=1,2$ and $5$. The hollow symbols are the numerical data of MFPs. The dotted, solid and dashed lines are the curves of $\hat{l}_k=A(T)\frac{\sin{k}}{\omega^2_k+\gamma}$ where $A(T)$ is the fitting parameter for $T=1,2$ and $5$ respectively. Here the fitting $A(T)$ is $A(T=1)=18.3$, $A(T=2)=15.0$ and $A(T=5)=12.0$. The calculations are performed on lattices with size $N=1024$.}
\end{figure}

\textbf{$\phi^4$ lattice.} The $\phi^4$ lattice has an on-site potential and exhibits normal heat conduction behavior \cite{Aoki2000pla,Hu1998pre,Hu2000pre}. According to definition, the renormalization coefficient $\alpha$ can be obtained as $\alpha=1$. The expressions of renormalized phonon frequency and its group velocity are
\begin{equation}
\hat{\omega}_k=\sqrt{\omega^2_k+\gamma},\,\,\hat{v}_k=\frac{\sin{k}}{\sqrt{\omega^2_k+\gamma}},
\end{equation}
where the renormalization coefficient $\gamma$ can be derived from a classical field approach as $\gamma\approx 1.230\, T^{2/3}$\cite{Boyanovsky2004}.

According to Eq. (\ref{conjecture}), the MFPs $\hat{l}_k$ of renormalized phonons of the $\phi^4$ lattice should follow a dependence on nonlinearity strength $\varepsilon$ and wave vector $k$ as
\begin{equation}\label{l-phi4}
\hat{l}_k\propto \frac{1}{\varepsilon}\frac{\sin{k}}{\omega^2_k+\gamma}
\end{equation}
where $\gamma\approx 1.230\cdot T^{2/3}$ is valid for temperatures not high enough\cite{Nianbei2013}. Since the heat conduction is normal for the $\phi^4$ lattice, we are able to test both the frequency and temperature dependence of MFPs here.

We first test the $k$ or frequency dependence of MFPs $\hat{l}_k$ as in Eq. (\ref{l-phi4}). By introducing a temperature-dependent but $k$ independent fitting parameter $A(T)\propto \varepsilon^{-1}$, it is easy to have $\hat{l}_k=A(T)\frac{\sin{k}}{\omega^2_k+\gamma}$. In Fig. \ref{fig:phi4-k}, the numerical calculated MFPs are plotted as hollow symbols at three different temperatures $T=1,2$ and $5$. These data are compared with $\hat{l}_k=A(T)\frac{\sin{k}}{\omega^2_k+\gamma}$ with $A(T=1)=18.3$, $A(T=2)=15.0$ and $A(T=5)=12.0$ plotted as the dotted, solid and dashed lines in Fig. \ref{fig:phi4-k}, respectively. It can be seen that the $k$ dependence of numerically calculated MFPs can be qualitatively described by the conjectured $k$ dependence of $\hat{l}_k$ of Eq. (\ref{l-phi4}). For $T=1$ and $2$, the peak positions of conjectured curves are left-shifted compared to the numerical data. At $T=5$, the peak positions are consistent between numerical data and the conjecture. The MFPs around $k\rightarrow 0$ are underestimated by the conjecture and this effect might be understood by noticing that the numerical calculated renormalized phonon frequencies $\hat{\omega}_k$ are smaller than what are predicted by the renormalized phonon theory\cite{Sha2014prb}.

\begin{figure}
\includegraphics[width=0.8\columnwidth]{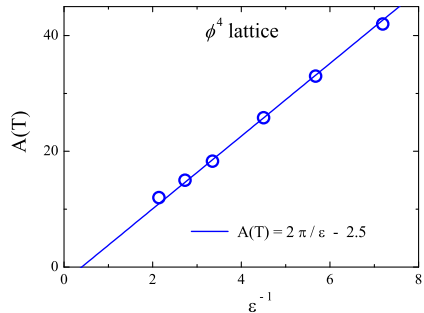}
\vspace{-0.5cm} \caption{\label{fig:phi4-epsilon} (color online). The fitting parameter $A(T)$ as the function of inverse nonlinearity strength $\varepsilon^{-1}$ for $\phi^4$ lattice. The $A(T)$ is fitted from the numerical calculated MFPs with the fitting formula of $\hat{l}_k=A(T)\frac{\sin{k}}{\omega^2_k+\gamma}$. The $\varepsilon$ is calculated in thermal equilibrium for a lattice with $N=100$. The solid line of $A(T)=2\pi/\varepsilon - 2.5$ is the guide for the eye.}
\end{figure}

The MFP $\hat{l}_k$ is also conjectured to be inversely proportional to the nonlinearity strength $\varepsilon$ for the $\phi^4$ lattice in Eq. (\ref{l-phi4}). The nonlinearity strength $\varepsilon$ can be expressed as $\varepsilon=\left<q^4_i/4\right>/(\left<(q_{i+1}-q_i)^2/2\right>+\left<q^4_i/4\right>)$
 by noticing the ensemble average is independent of atom index $i$. Unfortunately the $\varepsilon$ cannot be analytically calculated here. We can only numerically calculate the nonlinearity strength $\varepsilon$ at different temperatures at thermal equilibrium. In Fig. \ref{fig:phi4-epsilon}, the fitting prefactors $A(T)$ are plotted as the function of $\varepsilon^{-1}$ for temperatures ranging from $T=0.1$ to $T=5$. The linear dependence between $A(T)$ and $\varepsilon^{-1}$ verifies the dependence of nonlinearity strength for the MFP $\hat{l}_k$ conjectured as in Eq. (\ref{l-phi4}) for the $\phi^4$ lattice. Most interestingly, the slope of $2\pi$ of this dependence suggests the renormalized phonon's MFP $\hat{l}_k$ of the $\phi^4$ lattice can be written as
\begin{equation}
\hat{l}_k=\hat{v}_k\frac{1}{\epsilon}\frac{2\pi}{\hat{\omega}_k}
\end{equation}
where $2\pi/\hat{\omega}_k$ is nothing but one period of renormalized phonon with mode $k$!

In summary, we have systematically investigated the temperature and frequency dependence of MFPs of renormalized phonons in 1D nonlinear lattice with the newly developed tuning fork method. The conjecture made in the effective phonon theory is very verified suggesting the MFPs should be inversely linear proportional to the renormalized frequency and nonlinearity strength. This is different from the analysis of Umklapp phonon scattering theory where a MFP$\propto\omega^{-2}$ dependence is predicted \cite{Klemens1994}. But our work is consistent with the recent results numerically obtained for carbon nanotubes where the dependence of MFP$\propto\omega^{-1}$ is claimed\cite{Volz2015prb}. The current work reinforces the role of renormalized phonons during the heat transport process in strong nonlinear materials where the nonlinear effect cannot be ignored.

This work has been supported by the NSF China with grant No. 11334007 (N.L., B.L.), the NSF China with Grant No. 11205114 (N.L.), the Program for New Century Excellent Talents of the Ministry of Education of China with Grant No. NCET-12-0409 (N.L.) and the
Shanghai Rising-Star Program with grant No. 13QA1403600 (N.L.).

\bibliographystyle{apsrev4-1}

\end{document}